\documentclass{pasj00}

\begin{document}
\SetRunningHead{M.~S.~Tashiro et al.}{Hard X-ray Spectrum from Fornax~A West Lobe}
\Received{2008/08/16}
\Accepted{2008/09/11}

\title{Hard X-Ray Spectrum from West Lobe of Radio Galaxy Fornax~A Observed with Suzaku}




%
\author{%
  Makoto~S. \textsc{Tashiro}\altaffilmark{1},
  Naoki \textsc{Isobe}\altaffilmark{2},
  Hiromi \textsc{Seta}\altaffilmark{1},
  Keiko \textsc{Matsuta}\altaffilmark{3},
     \and
  Yuichi \textsc{Yaji}\altaffilmark{1}
}
\altaffiltext{1}{Saitama University, 255 Shimo-Okubo, Sakura, Saitama, 338-8570}
\email{tashiro@phy.saitama-u.ac.jp}
\altaffiltext{2}{RIKEN, 2-1 Hirosawa, Wako, Saitama, 351-0198} 
\altaffiltext{3}{JAXA/Institute of Space and Astronautical Science, 3-1-1 Yoshinodai, Sagamihara, Kanagawa, 229-8510}


\maketitle

\begin{abstract}
An observation of the West lobe of radio galaxy Fornax~A (NGC~1316) with Suzaku is reported.
Since \citet{feigelson95} and \citet{kaneda95} discovered the cosmic microwave background
boosted inverse-Comptonized (IC) X-rays from the radio lobe, the magnetic field and electron
energy density in the lobes have been estimated under the assumption that a single component
of the relativistic electrons generates both the IC X-rays and the synchrotron radio emission. However, electrons generating the observed IC X-rays in the 1 -- 10 keV band do not possess sufficient energy to radiate the observed synchrotron radio emission under the estimated magnetic field of a few $\mu$G. On the basis of observations made with Suzaku, we show in the present paper that a 0.7 -- 20 keV spectrum is well described by a single power-law model with an
energy index of 0.68 and a flux density of $0.12\pm0.01 \mu$Jy at 1 keV from the West lobe. The derived multiwavelength spectrum strongly suggests that a single electron energy distribution over a Lorentz factor $\gamma = $ 300 -- 90000 is responsible for generating both the X-ray and radio emissions. The derived physical quantities are not only consistent with those reported for the West lobe, but are also in very good agreement with those reported for the East lobe.
\end{abstract}

\section{Introduction}\label{sec:intro} 
Cosmic jets of active galactic nuclei (AGNs) are significant kinetic energy channels from supermassive blackholes to intergalactic space.
Some bulk flows release their entire kinetic energy into expanding radio lobes, which can be regarded as calorimeters for AGN outflows.
The injected kinetic energy is thought to be turned into internal energy of plasma and induced magnetic field energy in the lobes.
Although the energy densities of electrons and magnetic fields are often estimated assuming equipartition between the electron and magnetic fields, recent X-ray and radio imaging spectroscopy techniques have provided a way to resolve the two physical quantities without relying on this assumption.

From observations of the radio galaxy Fornax~A (NGC~1316) with ASCA and ROSAT, it was first discovered that the radio lobes generate inverse-Compton (IC) X-rays as well as a synchrotron radio emission. It was concluded that the IC X-rays were generated from the cosmic microwave background (CMB) photons (\cite{kaneda95}; \cite{feigelson95}), as has long been suspected (e.g. \cite{harris79}).
Following this discovery, through X-ray imaging spectroscopic measurements carried out by satellites in the past decade, including ASCA, ROSAT, Beppo-SAX, Chandra, and XMM-Newton, measurements of the electron energies of the lobes and the magnetic fields of a new level of precision have been achieved, in cooperation with measurements made using radio interferometers (e.g., \cite{brunetti01}; \cite{isobe02}; \cite{isobe05} hereafter I05; \cite{hardcastle02}; \cite{grandi03}; \cite{comastri03}; \cite{croston04}).

The derived electron energy densities often exceeded those of the magnetic fields by a factor of a few to several dozen in a number of the observed radio lobe objects (e.g. \cite{isobe02}; I05; \cite{croston05}). 
However, we note that there is still room for argument against this observational claim.
These evaluations are based on another assumption that the synchrotron and IC emissions are generated by relativistic electrons with a power law shaped energy spectrum; however, the validity of this law has not been strictly proven yet.
In the case of typical radio lobes, the radio synchrotron radiation is generated by electrons with $\gamma \sim (1-6) \times10^4$, whereas the observed IC X-rays in the range of 1 -- 10 keV are generated by electrons with $\gamma \sim (1-3) \times 10^3$.
In this regard, we need to examine the electron energy distribution in the range of $\gamma > 3 \times 10^3$ by observing IC X-rays with energies of over 10~keV.

We, therefore, used Suzaku to observe the hard X-ray extension of IC X-rays with energies of up to 20~keV to cover this observational gap. We expected to observe that the electrons producing the $\sim 20$ keV IC X-rays, would also be responsible for generating sub-100MHz synchrotron radio radiation.
Suzaku, which is equipped not only with a high-throughput X-ray telescope (XRT; \cite{xrt07}) and CCD cameras (XISs; \cite{xis07}) but also with the most sensitive hard X-ray spectrometer (HXD; \cite{hxd07}) currently available, is one of the most suitable instruments for the purpose of measuring the spectrum over 10~keV, the region corresponding to the electrons.

The structure of this paper is as follows. We describe the Suzaku observation and the results in \S~2 and \S~3. Following the presentation of the results, we evaluate the newly obtained X-ray spectra from the viewpoint of a multiwavelength spectrum and derive the physical quantities with respect to the lobe field in \S~4. Finally, we summarize these results in the last section.
In the Appendix, we present our analysis and results previously obtained by XMM-Newton in order to evaluate contaminant point sources in the West radio lobe region.
Throughout the present paper, we assume the distance to Fornax~A to be 18.6 Mpc (\cite{madore99}), on the basis of measurements made by the Hubble Space Telescope of Cepheid variables in NGC~1365, a galaxy that, along with Fornax~A, belongs to the Fornax Cluster. This distance gives an angle-to-physical size conversion ratio of 5.41 kpc~arcmin$^{-1}$ and a red shift of the rest frame  $z_{\rm rest}=0.00465$, while the measured red shift, including the proper motion, is reported to be $z_{\rm obs} = 0.005871$ (\cite{longhetti98}).

\section{Observations and data reduction} 
Suzaku observations of the Fornax~A West lobe were performed between 2005 December 23 20:51 UT and December 26 14:40 as part of the guest observation phase-1 of Suzaku (\cite{suzaku07}) (table~\ref{tab:obslog}).
Suzaku operated three sets of X-ray telescopes (\cite{xrt07}), each coupled to a focal-plane X-ray CCD camera known as an X-ray imaging spectrometer (XIS ; \cite{xis07}) covering the energy range of 0.2 -- 12 keV. Two (XIS0 and XIS3) of the XIS sensors use front-illuminated (FI) CCDs, while XIS1 utilizes a back-illuminated (BI) one, thus achieving an improved soft X-ray response. The hard X-ray detector (HXD; \cite{hxd07}; \cite{hxd2}) covers the 10 -- 70 keV energy band with Si PIN photo-diodes (HXD-PIN), and the 50 -- 600 keV range with GSO scintillation counters (HXD-GSO).
The HXD-PIN covers the energy band that is crucial to the evaluation of the electron energy distribution mentioned in \S~1. 
Since the optical axes of the XIS and the HXD are slightly offset ($\sim 3'$), 
we placed the lobe center at the center of the field of view of the HXD. 
Since the field of view of the HXD is so wide that it covers the host galaxy in the same field of view, we also performed offset observation pointing at the host galaxy at the aim point of XIS in order to evaluate the X-ray contamination from the host galaxy and the non-X-ray background (NXB) (table~\ref{tab:obslog}). 
The HXD and XIS were operated in nominal mode throughout the observations. 

The data retrieved from XIS were prepared with Processing Version 2.0.6.13, and reprocessed with the latest calibration database using the HEAdas 6.4.1 software package. 
We employed only the time regions whose geomagnetic cutoff rigidity (COR) was greater than 6 GV and removed data that suffered from telemetry saturation.
The XIS data was further screened using the following standard criteria: the XIS GRADE should be 0, 2, 3, 4, or 6, the time interval after the exit from the South Atlantic Anomaly should be longer than 436 s, and the object should be at least $5^\circ$ and $20^\circ$ above the dark and sunlit Earth rim. 
The HXD data was also screened using the following standard criteria: the time interval after the exit from the South Atlantic Anomaly should be longer than 500 s; and the object should be at least $5^\circ$ above both the dark and the sunlit Earth rim. 

\begin{table*}
\caption{Observation log}\label{tab:obslog}
\begin{center}
 \begin{tabular}{ccccccc}
 \hline \hline 
\multicolumn{3}{c}{Aim point (J2000)} & Start Date  &  Obs. ID   & $t_{\rm exp}$\footnotemark[$\dagger$]\\
     & R.A.    &  Dec.                &             &            & (ks)\\      
 \hline 
host galaxy\footnotemark[$*$]
            &\timeform{3h22m42.79s}&\timeform{37D12'29.88''} & 2006-12-22& 801015010  & 42.9\\ 
West lobe & \timeform{3h21m28.10s}&\timeform{37D7'38.08''} & 2006-12-25  & 801014010& 86.8\\ 
 \hline 
   \end{tabular} 
 \end{center}
\footnotemark[$\dagger$]: Net exposure time.\\
\footnotemark[$*$]:  Observation of the host galaxy is used only for background evaluation in the present paper.
\end{table*}

\section{Data Analysis and Results} 
\subsection{Point source and diffuse background reduction of XIS data}\label{sec:xis}
\begin{figure}
  \begin{center}
    \FigureFile(80mm,80mm){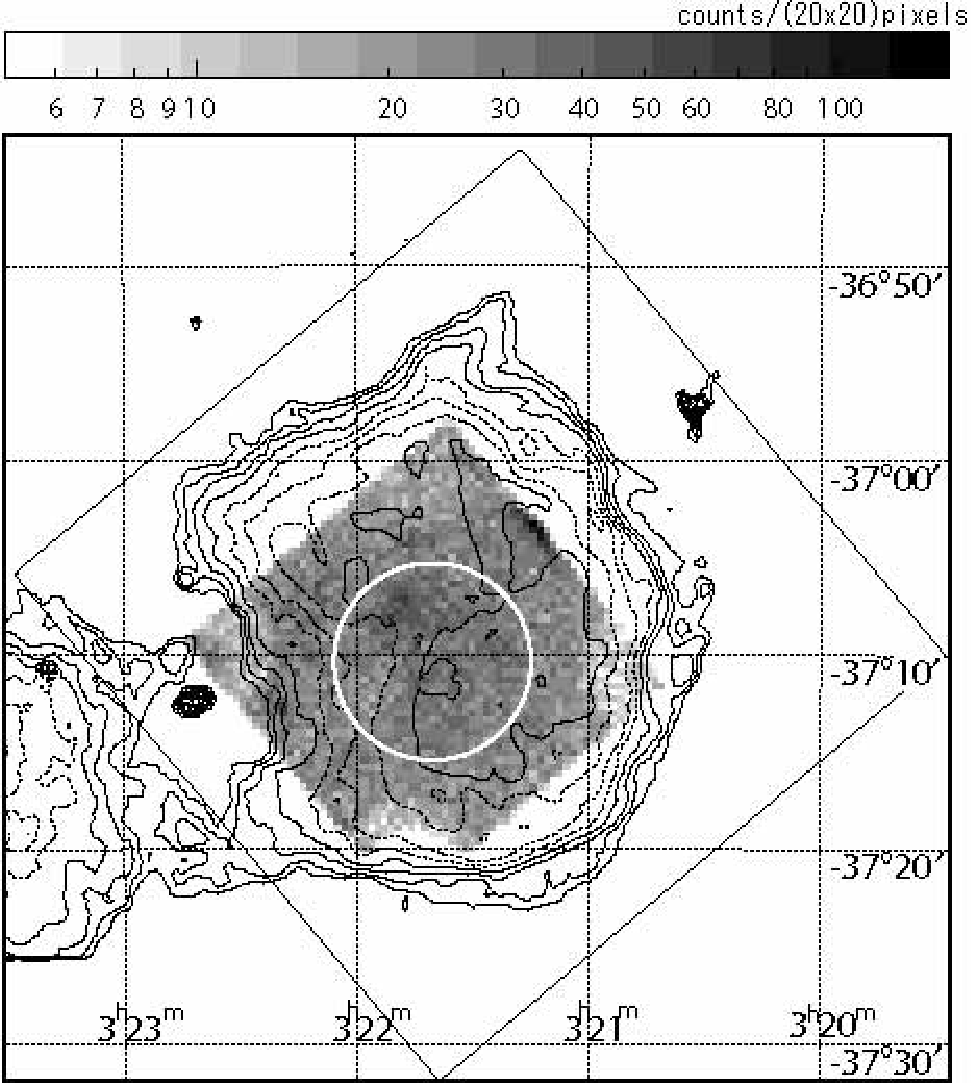}
  \end{center}
  \caption{The summed raw X-ray image obtained by XIS 0, 1, and 3 is shown in gray scale. The radio image is shown with overlaid contours. The FWHM of the field of view of the HXD/PIN is shown as a solid square. The white circle represents the XIS spectrum integration region (see text). The radio image was taken from \citet{fomalont89}.} \label{fig:image}
\end{figure}

We show a combined raw image obtained from the XISs in figure~\ref{fig:image}, for the fields of view. As indicated in figure~\ref{fig:image}, the entire XIS field of view of $18' \times 18'$ is covered with the extended emission. As previously reported by T01, the lobe X-ray emission extends up to $12'$ in radius, which fills the entire radio lobe region.
In the following analysis, we set the circle data acquisition region with a radius of $7.'25$, as indicated by a circle in figure~\ref{fig:image}. 
We confirmed that there was no significant time variation in the resultant lightcurve from the data acquisition region after performance of the data reduction procedures mentioned in \S~2. 

The XIS spectra obtained from the circular region were inevitably contaminated by both diffuse backgrounds (CXB and NXB) and point sources.
Since the entire field of view was covered with the source emission, we were unable to obtain a source-free region in the observed data.
The NXB spectrum of each XIS was created using a NXB database for the same region defined by the detector coordinates (DETX/Y). The database, supplied by the XIS team, is sorted by COR with a tool supplied by the XIS team ({\tt xisnxbgen}) in order to allow us to produce the model NXB at the same COR distribution. 
By means of this sorting process, the reproducibility of the NXB spectrum is improved to $< 4$ \% (\cite{tawa08}).
We also reproduced the CXB according to \citet{kushino02}.
The point source spectra we evaluated by utilizing XMM-Newton data as described in the Appendix. We thus estimated the contaminant sources, and subtracted the best fit model from the observed spectra; investigation of this is described in detail in \S~\ref{sec:jointfit}.

\subsection{Background reduction of HXD data}\label{sec:hxd}
The evaluated point source spectrum is too soft to have an effect on the hard X-ray spectrum. 
Therefore diffuse background subtraction is the key in the HXD data reduction procedure. Here we employed model backgrounds of NXB and CXB for the HXD data.
The model background for the NXB was calculated as simulated photon data, whose model parameters were continuously tuned using the Earth occultation data (\cite{fukazawa09}).
Before adopting the model background --- called ``tuned-''NXB, we evaluated the count rate reproducibility of tuned-NXB by comparing it with those of the Earth occultation data for during our observation.
Using {\tt Xselect}, we extracted the lightcurve from the time region for which the field of view of the HXD/PIN was fully covered by the Earth ({\tt ELV < - 5 deg}) from the observation data and the tuned-NXB, to allow comparison of the ``lightcurves'' of the NXB.
The count rate ratio in the 15 -- 40 keV region, where (observed data)/(model NXB) = $0.982 \pm 0.012$, is nearly equal to unity, although the discrepancy corresponds to within $2 \sigma$ of the model reproducibility (\cite{fukazawa09}).

\begin{figure}
  \begin{center}
    \FigureFile(80mm,80mm){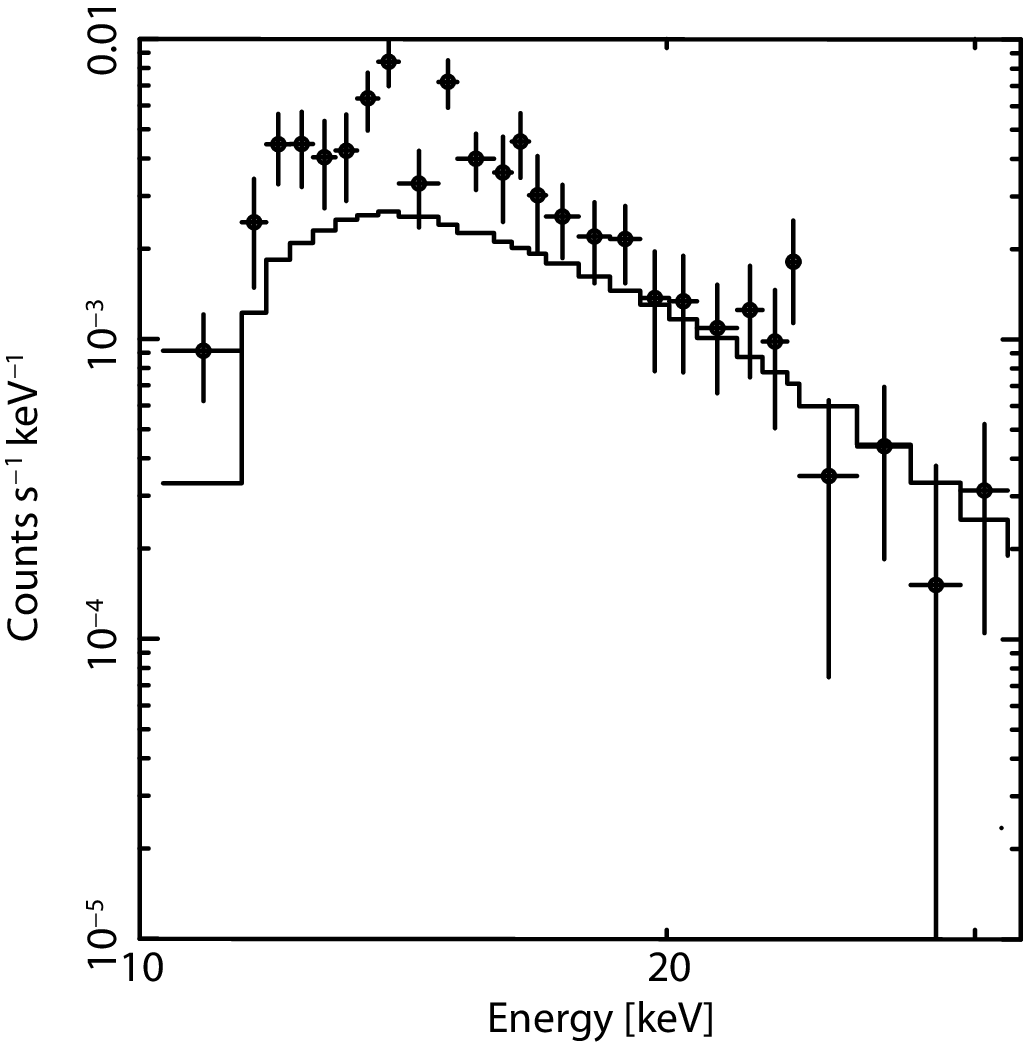}
  \end{center}
  \caption{HXD/PIN spectrum after rescaled model NXB subtraction. The residual data points (crosses) exceed those in the model CXB (histogram) significantly in the range of 10 to 20 keV}\label{fig:hxdspec}

\end{figure}

As a reference, we rescaled the model NXB by a factor of 0.982 and subtracted it from the spectrum in order to compare it with the model CXB in figure~\ref{fig:hxdspec}.
According to \citet{bolt87}, here we adopted a fixed CXB component for the HXD/PIN spectrum described with:
\tiny
\begin{displaymath}
F(E) = (1.05 \times 10^{-3}) \left( \frac{E}{\rm 1 keV} \right)^{-0.29} 
	\exp \left[ \frac{E}{\rm 40 keV} \right] 
	\;\; {\rm keV~s}^{-1}{\rm cm}^{-1}{\rm keV}^{-1}.
\end{displaymath}
\normalsize
The figure shows that the intensity of the PIN spectrum from the West lobe region significantly exceeds the expected CXB spectrum.
In the 12 -- 25~keV band, both the rescaled NXB spectrum and the CXB model subtracted spectrum exhibit $(1.41 \pm 0.22) \times 10^{-2}$ count~s$^{-1}$ with the significance level of $6.43 \sigma$. 
Even if we adopt the nominal tuned-NXB as the most conservative case, the background subtracted spectrum exhibits $(0.75 \pm 0.22) \times 10^{-2}$ count~s$^{-1}$ with a significance level of $3.38 \sigma$.

In order to check the consistency of the data, we examined the consistency of the background estimation and extension of the West lobe emission, using the offset observation data. 
Although the offset observation was directed at the host galaxy, the field of view still covered a part of the lobe emission, at a fraction of the full effective area (see figure~\ref{fig:image}).
Assuming that the lobe emission extends over $12'$ in radius (T01), we calculated the effective area for the West lobe from the lobe observation attitude ($A_{\rm lobe}$) and the offset (host galaxy) observation attitude ($A_{\rm off}$). The ratio of these effective areas was thus evaluated to be $A_{\rm _lobe}/ A_{\rm_off} =  1.41$.
If the West lobe is the dominant source for the HXD, the background-subtracted count rates of (A) (lobe emission) $-$ (model NXB + CXB) and that of (B) (lobe emission) $-$ (host galaxy data) are expected to be proportional to the respective effective areas, $A_{\rm off}$ and $A_{\rm lobe}$, since no or very weak hard X-ray emission from the host galaxy has been repeatedly reported by \citet{kaneda95}, \citet{iyomoto98}, T01, and \cite{kim03}.
In fact, we confirmed that the ratio of (A)/ (B) = $1.02 \pm 0.26 ^{+1.20}_{-0.00}$ is consistent with the effective area ratio as calculated above, where the former and the latter errors represent the statistical and the systematic errors between the tuned and rescaled NXB models.

\subsection{Wide-band X-ray spectra from XIS and HXD}\label{sec:jointfit}
\begin{figure}
  \begin{center}
    \FigureFile(80mm,80mm){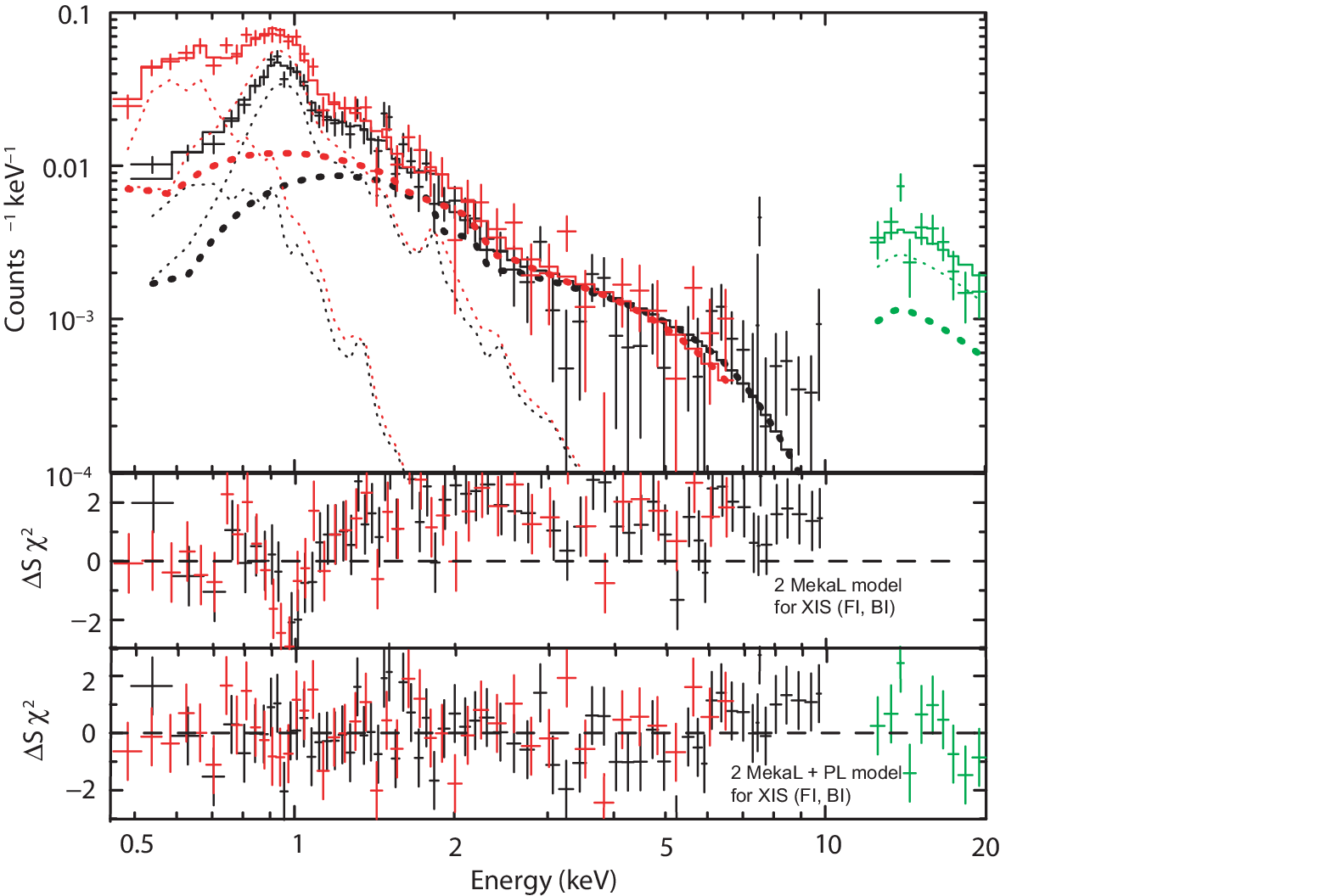}
  \end{center}
  \caption{
Joint X-ray spectra from XIS-FI (from the XIS-1; in red), XIS-BI (from the XIS-0 and 3 summed; in black), and HXD-PIN (in green). The employed models comprise two thin thermal plasma emissions (thin dotted lines in red and black), a power law component for the lobe emission (dashed lines), and the CXB model for the HXD (thin dotted line in green), which are shown in the total model spectra with solid lines. The middle and bottom panel show residuals of XIS data for the two thermal emission models only and residuals of both XIS and HXD data for the total model including the additional power law component, respectively. The models and the obtained best fit parameters are summarized in table~\ref{tab:fitlog}}\label{fig:xishxdspec}
\end{figure}

We obtained the background-subtracted spectra from the XISs and HXD as shown in figure~\ref{fig:xishxdspec}.
We employed the response matrices for the XIS calculated for an integration circle of $7'.25$ radius, using {\tt xisrmf}. 
The XIS {\it arf} (ancillary response file) was simulated with {\tt xissimarfgen} (\cite{ishisaki07}), assuming the West lobe had a flat rightness distribution with a radius of 12' (T01)"
As for the HXD/PIN, we employed the response matrices {\tt ae\_hxd\_pinhxnome3\_20080129.rsp} generated with {\it arf}, assuming the same diffuse emission.
Before evaluating the hard X-rays emitted from the West lobe as detected by HXD/PIN, we examined two thin thermal plasma emission components (T01; I05) in the XIS spectra and found an unacceptable fit with $\chi^2 /$d.o.f = 362 / 109.
As can be seen in the middle panel of figure~\ref{fig:xishxdspec}, the XIS spectrum requires an additional hard component.

Then, we added a power law model for the expected non-thermal component to perform an XIS-HXD/PIN joint spectrum fitting.
The derived best fit parameters are summarized in middle column of table~\ref{tab:fitlog}.
The additional power law successfully fit the 0.5 --- 20 keV spectrum with the energy index of $0.81\pm0.22$, which is consistent with that measured in the radio band ($\alpha =0.68\pm0.10$; I05).

Finally, we introduced the spectral slop determined by radio observation, and evaluated the XIS-HXD/PIN spectrum. The results are shown in the top and bottom panels of figure~\ref{fig:xishxdspec}, and the derived best fit parameters are summarized in the right column of table~\ref{tab:fitlog}
Thus, we obtained the best fit X-ray flux density at 1~keV of $0.12\pm0.01 \mu$Jy, with an energy index of 0.68 (fixed), and confirmed that the newly obtained fitting parameters are consistent with those presented in previous reports (\cite{kaneda95}; T01)
Note that we showed only the nominal tuned-NXB subtracted HXD/PIN spectrum above as the most conservative case. However we also confirmed that there was no significant difference in the obtained best fit parameters whether we employed the nominal or rescaled NXB (\S~\ref{sec:hxd}).

\begin{table*}
\caption{Summary of spectral fitting}\label{tab:fitlog}
\begin{center}
 \begin{tabular}{lcc}
 \hline \hline 
   model components               & model I\footnotemark[$\dagger$] 
                                  & model II\footnotemark[$\ddagger$] \\
 \hline 
  (a) Column density (cm$^{-2}$)  &  $2.06\times10^{20}$ (fix) & $\leftarrow$ \\
  (b) $kT$         (keV)          &  $0.197^{+0.017}_{-0.014}$ & $\leftarrow$ \\ 
  (b) Number density (cm$^{-2}$)  &  0.001 (fix)               & $\leftarrow$ \\ 
  (b) Abundance   (solar)         &  0.3 (fix)                 & $\leftarrow$ \\ 
  (b) Red shift                   &  0.005871 (fix)            & $\leftarrow$ \\ 
  (b) Normalization               & $(1.35\pm0.02)\times10^{-3}$ &$\leftarrow$ \\
  (c) $kT$         (keV)          & $0.81\pm0.04$              & $\leftarrow$ \\
  (c) Number density (cm$^{-3}$)  &  0.002 (fix)               & $\leftarrow$ \\
  (c) Abundance   (solar)         &  0.3 (fix)                 & $\leftarrow$ \\
  (c) Red shift                   &  0.005871 (fix)            & $\leftarrow$ \\
  (c) Normalization               &   $(7.9\pm0.7)\times10^{-4}$ & $\leftarrow$ \\
  (d) Flux density ($\mu$Jy at 1 keV) &  $0.13\pm0.03$             & $0.12\pm0.01$ \\
  (d) Energy index ($\alpha$)     &  $0.81 \pm 0.22$       & 0.68 (fix) \\
  $\chi^2$ / d.o.f.               &  138/116               & 139/117\\
 \hline 
   \end{tabular} 
\end{center}
\footnotemark[$\dagger$]: Model components for XIS spectrum; (a) Galactic absorption 
$\times$ {(b) + (c) thin thermal plasma emissions from IGM (MEKAL)} + (d) power law (lobe emission; energy index is a free parameter); HXD spectrum; (d) power law (simultaneously fit with XIS spectra) + CXB (fixed: see text).\\
\footnotemark[$\ddagger$]: The same model as model I is employed but the energy index is fixed to the value measured in the radio band.
\end{table*}

\section{Discussion} 
\begin{figure}
  \begin{center}
    \FigureFile(80mm,80mm){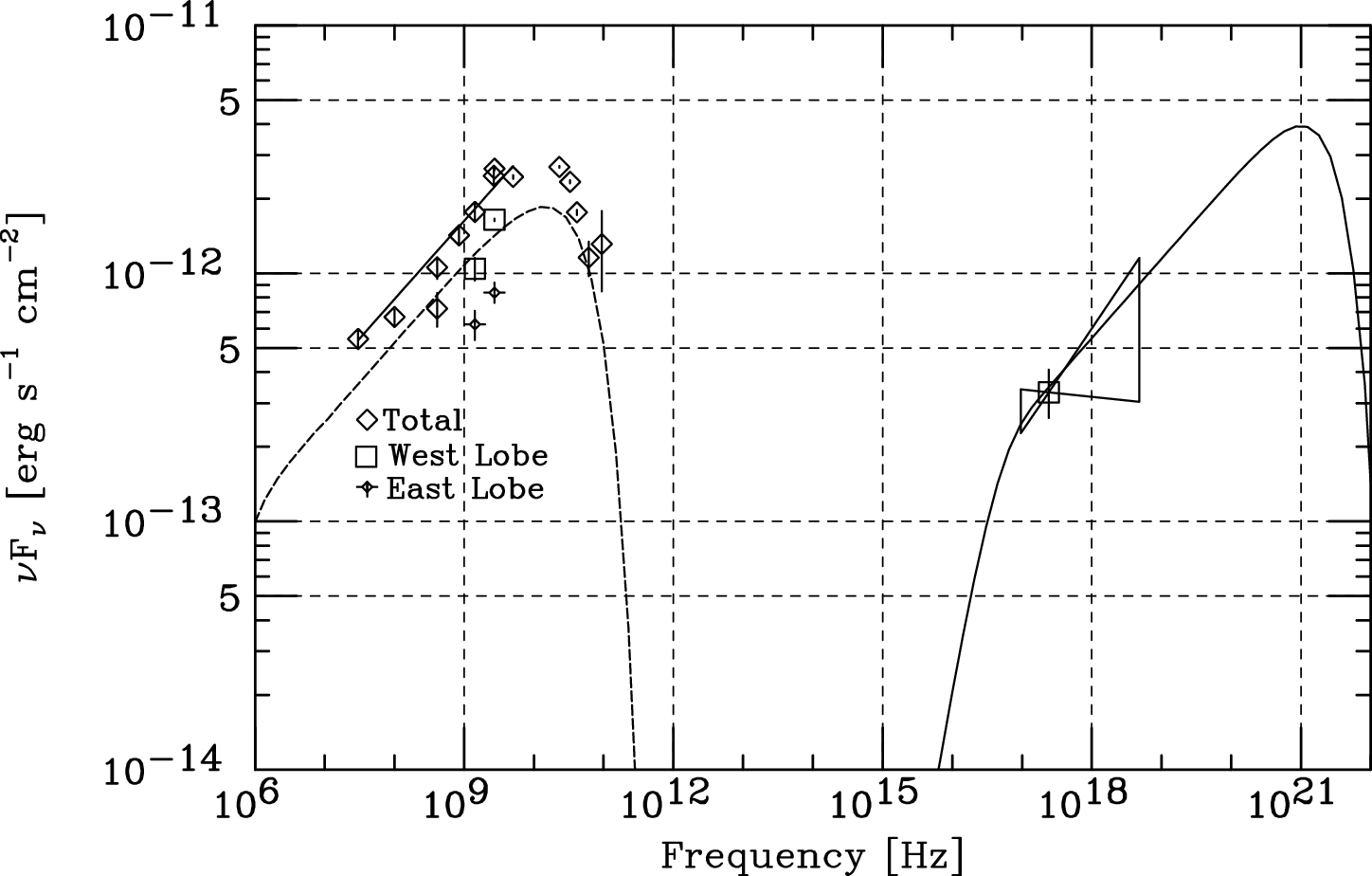}
  \end{center}
  \caption{(A) Spectral energy distribution from the West lobe (squares), East lobe (smaller diamonds with crosses), and total (larger diamonds). Derived models for the synchrotron emission (dashed line) and the CMB-boosted IC emissions (solid line) are shown with lines (see text). The 29.9 and 100 MHz data points are taken from \citet{finlay73}, the 408 MHz data points from \citet{robertson73}, the 843 MHz data points from \citet{jones92}, the 1.4 GHz data points from \citet{ekers83}, the 2.7 GHz data points from \citet{ekers69}, \citet{shimmins79} and \citet{kuehr81}, and the 5.0~GHz data point from \citet{kuehr81}.}\label{fig:sed}
\end{figure}

We succeeded in observing non-thermal X-rays from the Fornax~A West lobe with the HXD/PIN instrument onboard Suzaku.
Having assumed a hard X-ray emission extending over the lobe with a radius of $12'$, we confirmed that the measured hard X-ray spectral slope and flux are in good agreement with the extrapolation of the diffuse non-thermal component observed with XIS below 10~keV.
The non-thermal X-rays can be described sufficiently well by a single power law model over the entire range from 0.5 up to 16 keV (with the nominal NXB) or to 20 (with the rescaled NXB) keV, and the spectral slope is consistent with that observed in the radio spectrum.
Employing the derived power law component for the diffuse hard X-rays, we plotted the spectral energy distribution (SED) together with the radio spectra in figure~\ref{fig:sed}.

An X-ray spectrum following a single power law distribution, which does not smoothly connect to the synchrotron component, requires another origin.
The IC process is the most promising origin for the observed hard X-rays reported so far.
Taking the size of and the distance from the host galaxy into account, we found that the photon energy density of the CMB at the source rest frame ($4.3\times 10^{-13}$erg~cm$^{-3}$) exceeds both emissions from the host galaxy emission ($\sim$ a few $\times 10^{-15}$erg~cm$^{-3}$) and synchrotron emission from the lobe ($\sim 6 \times 10^{-17}$erg~cm$^{-3}$) by two orders of magnitude (e.g. \cite{rband07} and references in the caption of figure~\ref{fig:sed}).

In accordance with \citet{harris79}, we derived the estimated energy density of the electrons and the magnetic field energy density to be $u_{\rm e} = (5.1 \pm 1.0) \times 10^{-13}$erg~cm$^{-3}$
 and $u_{\rm m} = (0.67 \pm 0.08)\eta^{-1} \times 10^{-13}$erg~cm$^{-3}$, respectively, where $\eta$ is the filling factor of magnetic field against the electron spatial distribution.
Here, we assumed an electron Lorentz factor of $\gamma = 300 - 90000$, as we describe below, and included the systematic errors arising from the ambiguity of the reported radio flux and slope according to \S~4 in I05.
Note that the difference in brightness profiles between radio and X-rays is discussed with ASCA (T01). The displacement is thought to originate from partial displacement between the magnetic field and the electron distribution. T01 examined it and showed the magnetic field filling factor could be $\eta = 0.95$ or less, with the electrons filling the entire volume homogeneously.

Adopting the above values, we calculated the CMB boosted IC spectrum and plotted the results in the obtained SED in figure~\ref{fig:sed}.
Here, we assumed an electron differential energy spectrum with a single power law distribution of
\begin{displaymath}
N_{\gamma} = 2.0 \times 10^{-6} \gamma^{-2.36} \;\; {\rm electrons~cm}^{-3},
\end{displaymath}
Here, we employed the lower boundary electron energy distribution of $\gamma = 300$ to describe the lower energy end of the detected X-rays, although we have no evidence of a lower energy cutoff of the electron energy distribution.
Similarly, we set the upper boundary of $\gamma=90000$ in accordance with the higher end of the radio observation of the total lobe observations.
The calculation program was written to use the SSC equations presented in \citet{kataoka00}, to which we added the CMB/IC spectrum, and it can be seen that it provides a good description of the obtained multiwavelength data.

The electron Lorentz factor of $\gamma = 300 - 4000$ (or 4500) is required by the CMB/IC X-ray spectrum in the range of 0.5 -- 16 (or 20) keV, where we employed the nominal (or rescaled) NXB model for the HXD/PIN. At the same time, adopting the derived magnetic field, the observed synchrotron radiation spectrum requires a Lorentz factor of $\gamma > 4200$.
Thus, we conclude that a single electron energy distribution can naturally explain the two independent measurements in radio and X-ray bands.
The derived parameters are summarized in table~\ref{tab:param}.
We also note that the derived electron and magnetic field energy densities are not only consistent with those determined in previous reports (T01) but are also in very good agreement with those determined for the East lobe (I05).
	
\begin{table}
 \caption{Derived physical quantities}\label{tab:param}
 \begin{tabular}{cccc}
 \hline \hline 
                       & West Lobe    & East Lobe\footnotemark[$\dagger$] & unit  \\
 \hline 
  (1) $B_{\rm eq}$     & 1.59          & 1.55                   & $\mu$G \\
  (2) $B_{\rm IC}$     & $1.3\pm0.1$   & $1.23^{+0.08}_{-0.06}$ & $\mu$G \\
  (3) $u_{\rm m}$      & $0.67\pm0.08$ & $0.60^{+0.08}_{-0.06}$ & $10^{-13}$ erg~cm$^{-3}$  \\
  (4) $u_{\rm e}$      & $5.0\pm1.0$   & $5.0^{+1.1}_{-1.0}$    & $10^{-13}$ 
                                                             erg~cm$^{-3}$   \\
  $u_{\rm e}/u_{\rm m}$ &  $7.5 \pm 0.6$ & $5.0^{+1.1}_{-1.0}$  &  \\
 \hline 
   \end{tabular} 
\footnotemark[$\dagger$]: Taken from I05 for reference.\\
(1) Estimated magnetic field assuming electron-magnetic field energy equipartition.\\
(2)---(4) Magnetic field strength, magnetic field energy density, and electron energy density as derived from the observed CMB/IC X-rays and the synchrotron radio waves, respectively (see the text).
\end{table}

\section{Summary} 
Using Suzaku/HXD, we succeeded in detecting X-ray photons with energies up to 16 -- 20~keV from the West lobe of Fornax~A. A single power law provides a good description of the obtained 0.5 -- 20 keV spectrum.
The obtained properties of the single-component non-thermal emissions in the X-ray and radio bands strongly support the hypothesis that the IC X-rays are generated by electrons generating a synchrotron radio emission whose Lorentz factor ranges between $300$ and $90000$.
Although we assumed that the hard X-ray brightness profile was the same as that observed with ASCA in the 1 -- 10 keV range, the validity of this assumption still needs be confirmed from additional HXD mapping or hard X-ray imaging with next-generation satellites.

\bigskip	
The authors thank the {\it Suzaku} team for their great efforts in the observations and the calibration procedures.

\appendix
\section*{Point source detection with XMM-Newton}
\subsection*{XMM-Newton Data Reduction}
\begin{figure}
  \begin{center}
    \FigureFile(80mm,80mm){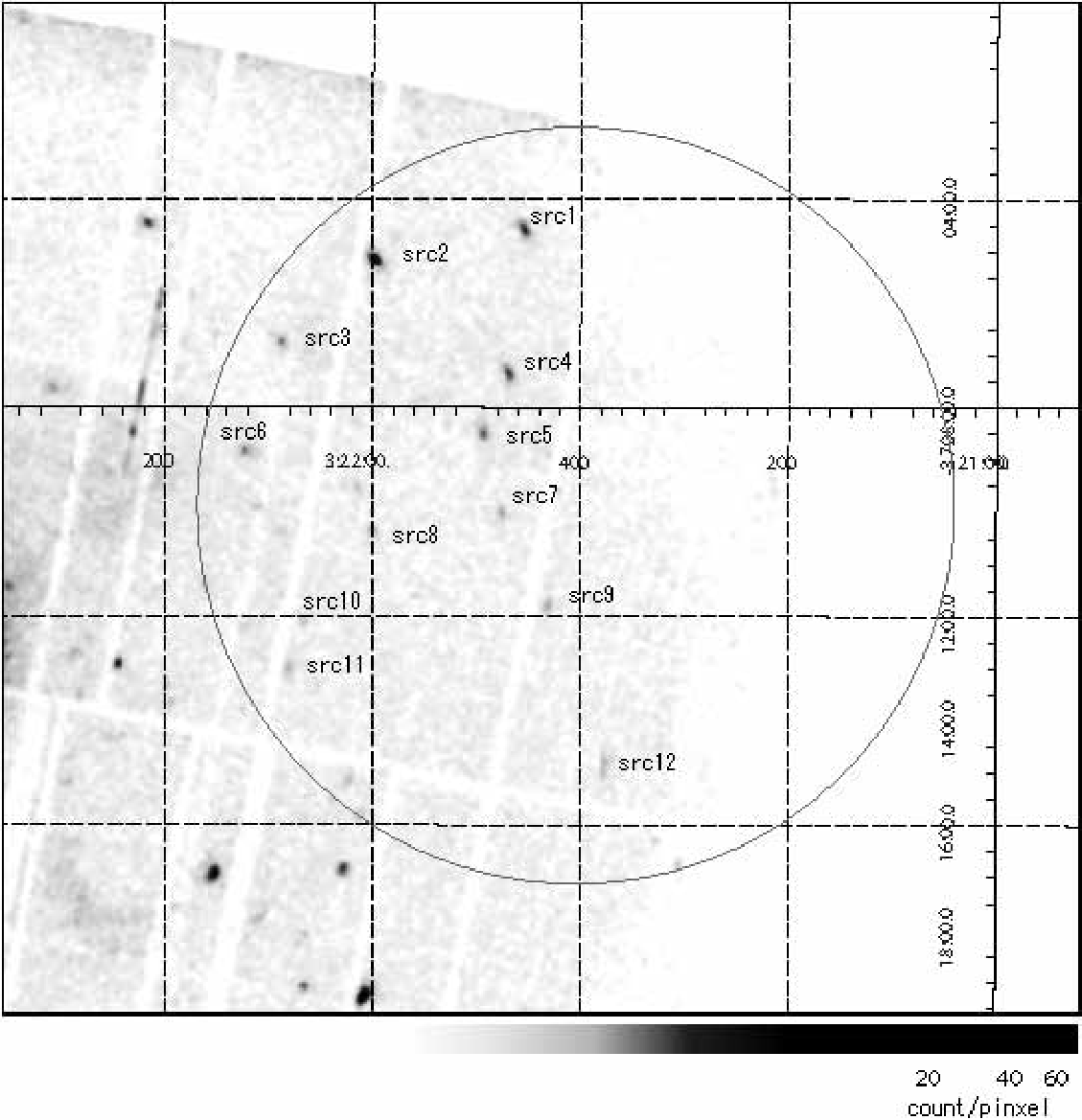}
  \end{center}
  \caption{The X-ray image obtained by XMM-Newton MOS1 in grayscale is shown with the XIS spectrum integration region indicated by a solid circle (figure~\ref{fig:image}). The evaluated point source positions are marked ``src 1 -- 12''.}\label{fig:xmm-image}
\end{figure}

We examined the XMM-Newton data in order to evaluate possible contaminant point sources within the lobe. The observation (observation ID = 0302780101) was performed on August 11, 2005, although the field of view of this observation did not cover the entire data acquisition region adopted for the XIS (figure~\ref{fig:image}).
We employed Science Analysis System (SAS) version 7.1.2 to reduce the volume of data.
All the data were reprocessed by {\tt emchain} or {\tt epchain} on the basis of Current Calibration Files (CCFs), the latest version of which were created at the beginning of April 2007. 
We removed high-background periods with {\tt PATTERN == 0} and {\tt \#XMMEA\_EM/P} in 10 -- 15 keV, utilizing a point-source-removed lightcurve (\cite{read03}).
We thus obtained good exposures of 74.7, 80.7 and 51.3 ks, for MOS1, MOS2 and PN, respectively.
The obtained X-ray image is shown in figure~\ref{fig:xmm-image}. 
For the spectral analysis, we selected MOS events with {\tt PATTERN $\le 12$}, {\tt \#XMMEA\_EM} and {\tt FLAG = 0}, 
and PN ones with {\tt PATTERN $\le 4$}, {\tt \#XMMEA\_EP} and {\tt FLAG = 0}.

\subsection*{X-ray spectra of contaminating sources}
We found twelve point sources within the XIS data acquisition circle. We accumulated counts of photons that fell within $36''$ from the center of each point source position and subtracted background accumulated from each concentric ring region with the inner and outer radii of $72''$ and $180''$, respectively.
Response files for each spectrum, {\it rmf} and {\it arf}, were generated with {\tt rmfgen} and {\tt arfgen}.
In this spectral analysis, we used summed spectra obtained with MOS1 and MOS2 for all but two cases. Since the source 3 fell out of MOS1, we employed PN + MOS2 for this source. We used only PN for source 6 for the limited statistics of the MOS1 and MOS2 data.
All spectra except that of source 2 were well described with the power law model; source 2 required an additional power law model.
The employed model and derived best fit parameters are summarized in table~\ref{tab:xmm-fit}. 
Each point source, exhibiting 0.5 -- 10 keV flux 
of $(0.5 - 3.2) \times 10^{-14}$ erg~s$^{-1}$cm$^{-2}$, very faint in comparison with the total lobe flux (\S~\ref{sec:jointfit}).

\begin{table*}
\caption{Summary of XMM-Newton spectra of contaminating point sources} 
\label{tab:xmm-fit}
\begin{tabular}{llllllll}
\hline 
Name       & (Ra, Dec)	         & Model           &	$N{\rm H}$                  & $\Gamma_1$              & $\Gamma_{2}$           
	     	     & $F_{\rm 0.5-10~keV}$ \footnotemark[$\dagger$]                 & $\chi^2/{\rm d.o.f}$\\   
           &         & 	       		             & ($10^{20}$cm$^{-2}$)           &                         &                        
	   	     & ($10^{-14}$ erg~s$^{-1}$cm$^{-2}$)            &                     \\ 
\hline\hline 
src 1  & (50.4388, -37.0767) & PL                & $2.06$  \footnotemark[$*$]    & $2.18_{-0.27}^{+0.29}$  & --                     
       	     	     & 3.0                           & $40.8 / 34$        \\
src 2  & (50.4989, -37.0859) & PL + PL           & $2.06$  \footnotemark[$*$]    & $6.64_{-1.28}^{+1.73}$  & $1.83_{-0.48}^{+0.45}$ 
       	     	     & 2.9  \footnotemark[$\dagger$] & $58.3 / 44$        \\
src 3  & (50.5363, -37.1123) & PL                & $21.0~(<54.9)$                & $2.78_{-1.41}^{+2.04}$  & --                     
       	     	     & 1.4                           & $26.9 / 25$        \\
src 4  & (50.4448, -37.1226) & PL                & $2.06$  \footnotemark[$*$]    & $1.80_{-0.34}^{+0.35}$  & --                     
            	     & 3.2                           & $31.1 / 35$        \\ 
src 5  & (50.4552, -37.1420) & PL                & $30_{-23}^{+50}$              & $5.2_{-2.3}^{+4.2}$     & --                     
       	     	     & 0.9                           & $21.5 / 26$        \\
src 6  & (50.5514, -37.1467) & PL                & $16~(<75)$                    & $2.3_{-2.0}^{+4.2}$     & --                     
       	     	     & 1.1                           & $4.8  / 11$        \\
src 7  & (50.4482, -37.1669) & PL                & $2.06$  \footnotemark[$*$]    & $1.84_{-0.82}^{+0.88}$  & --                     
       	     	     & 1.1                           & $23.5 / 22$        \\
src 8  & (50.4999, -37.1727) & PL                & $2.06$  \footnotemark[$*$]    & $2.5_{-1.0}^{+1.1}$     & --                     
       	     	     & 0.5                           & $18.0 / 14$        \\
src 9  & (50.4230, -37.1967) & PL                & $13~(<52)$                    & $1.6_{-1.0}^{+1.9}$     & --                     
       	     	     & 2.4                           & $19.7 / 21$        \\
src 10 & (50.5283, -37.2017) & PL                & $2.06$  \footnotemark[$*$]    & $2.3_{-1.1}^{+1.3}$     & --                     
       	     	     & 0.5                           & $6.8  / 11$        \\
src 11 & (50.5340, -37.2165) & PL                & $2.06$  \footnotemark[$*$]    & $1.7_{-1.0}^{+1.2}$     & --                     
       	     	     & 0.8                           & $7.4  / 14$        \\ 
src 12 & (50.4071, -37.2482) & PL                & $2.06$  \footnotemark[$*$]    & $1.78_{-0.53}^{+0.52}$  & --                     
       	     	     & 2.4                           & $21.8 / 23$        \\
\hline 
\end{tabular}\\
\footnotemark[$*$]       fixed at the Galactic value\\
\footnotemark[$\dagger$] without absorption correction\\ 
\footnotemark[$\sharp$] sum flux of the two PL components 
\end{table*}

\subsection*{Contribution to the XIS spectra}
\begin{figure}
  \begin{center}
    \FigureFile(80mm,80mm){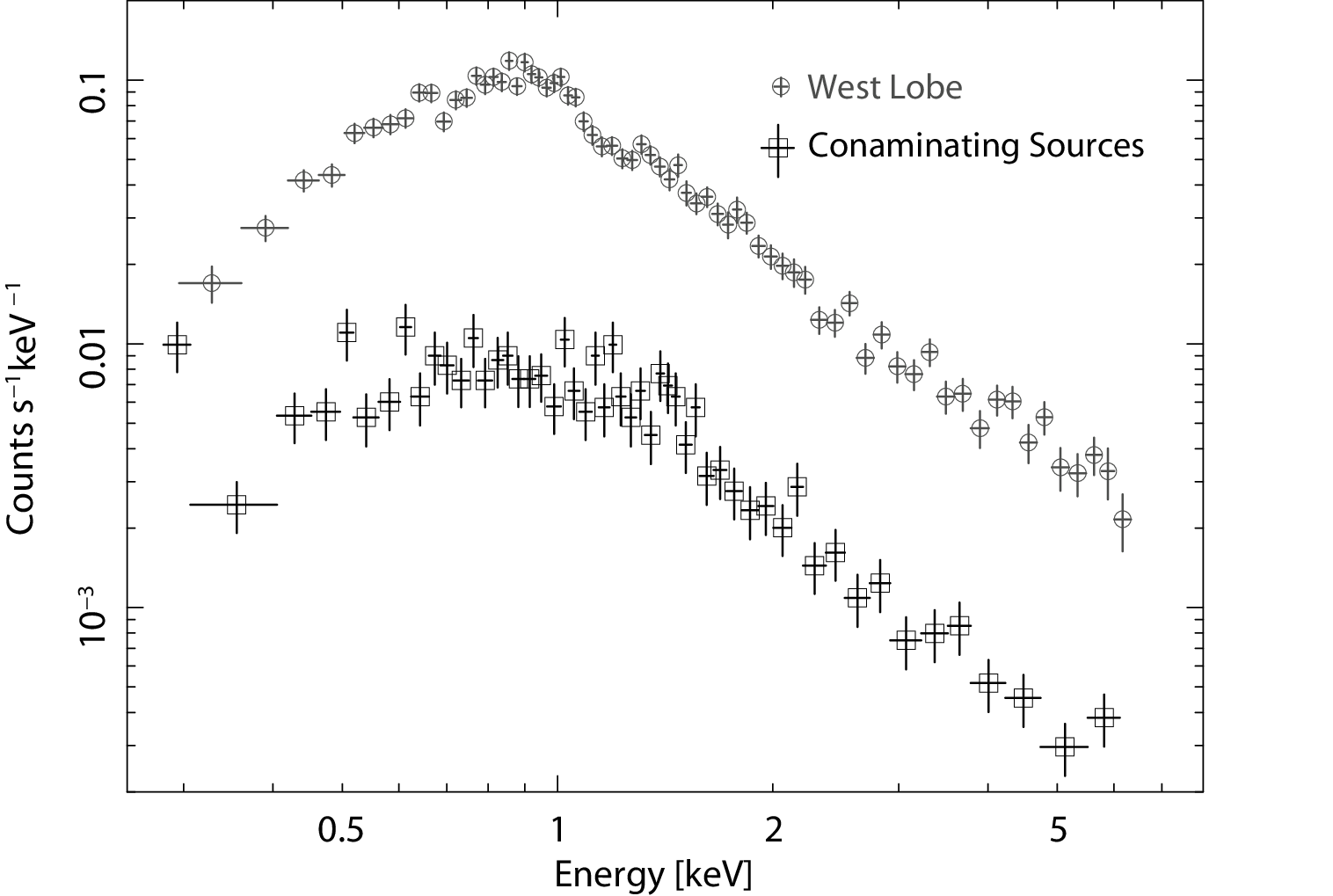}
  \end{center}
  \caption{Total X-ray spectrum of contaminant point sources evaluated with XMM-Newton data. The model and best fit parameters for each point source are summarized in table~\ref{tab:xmm-fit}. The NXB subtracted lobe spectrum obtained with XIS-1 is also presented in gray for reference.}\label{fig:xmmspec}
\end{figure}

We estimated the contribution of the twelve point sources to the XIS lobe spectrum integration region (\S~\ref{sec:xis}) using the following procedures.
First, we generated XIS {\it arfs} for each point source using {\tt xissimarfgen} with the XIS integration radius of $7.'25$, as we employed in \S~\ref{sec:jointfit}.
Then we simulated each point source spectrum, adopting the best fit parameters described above with the integration time of the XIS observation.
Finally, we added the simulated twelve spectra, as shown in figure~\ref{fig:xmmspec} for comparison with the NXB-subtracted XIS lobe spectrum.
We can see that the XIS spectrum, containing the West lobe emission and CXB, exceeds the total of the twelve point source spectra by an order of magnitude.
Since the XMM-Newton field of view covers about a half of the XIS integration region, we might have underestimated the contamination by a factor of a few. 
However, the evaluated point sources were so soft and the summed flux was so small that it is not likely to have affected the hard X-ray emission from the West lobe significantly.

\end{document}